\documentclass[aps,pre,twocolumn,showpacs,groupedaddress]{revtex4}

\usepackage{graphicx}
\usepackage{dcolumn}
\usepackage{bm}
\usepackage{amssymb}

\begin{document}


\title{Anisotropic surface tension of buckled fluid membrane}
\author{Hiroshi Noguchi}
\email[]{noguchi@issp.u-tokyo.ac.jp}
\affiliation{
Institute for Solid State Physics, University of Tokyo,
 Kashiwa, Chiba 277-8581, Japan}

\date{\today}

\begin{abstract}
Thin solid sheets and fluid membranes exhibit buckling
under lateral compression.
Here, it is revealed that buckled fluid membranes have
 anisotropic mechanical surface tension
contrary to solid sheets. 
Surprisingly, the surface tension perpendicular to the buckling direction
shows stronger dependence on the projected area than that parallel to it.
Our theoretical predictions are supported by
numerical simulations of a meshless membrane model.
This anisotropic tension can be used to measure the membrane bending rigidity.
It is also found phase synchronization occurs between multilayered buckled membranes.
\end{abstract}

\maketitle

\section{Introduction}

Buckling and crumpling of thin solid sheets and strips \cite{witt07,nels04}
are commonly seen in our daily life (e.g., for fruits \cite{yin08}, paper \cite{domo03,vlie06a}, and polyster film \cite{poci08}) 
as well as on the molecular scale (e.g., for viruses \cite{lidm03} and atomistic graphene sheets \cite{pere10}).
The shapes produced by the buckling of elastic sheets have fascinated many physicists.
In 1691,
J. Bernoulli proposed the problem of a simple bent beam or rod ``elastica'',
which led to the development of the calculus of variation and the elliptic function \cite{true83}.
The curve on a plane for minimum bending energy $\int^L c^2 ds$ with constant total length $L$
is expressed by elliptic functions, where $c$ is the curvature and $s$ is the arc length.
The theory of elastica was recently extended to twisted rods to describe the 
shape of supercoiled DNAs \cite{tsur86}
 and it has been employed to draw smooth surfaces in computer graphics \cite{bruc01}.

The buckling has been intensively investigated
for Langmuir monolayers on air-water interface \cite{miln89,lipp96,hatt03,zhan07,baou08,poci08a}.
The buckling develops to collapse or fold of the monolayers into water.
For a fluid membrane, the balance between gravity and membrane bending energy 
determines the buckling wavelength \cite{miln89}.
Recently, buckling transition was also observed in a fluid bilayer membrane in simulations
\cite{stec04,otte05,nogu06}.
For the bilayer membrane, the effects of gravitation and membrane dissolution to the supporting water
are negligible.
Thus, the bilayer membrane is a simpler system so it is suitable to
study the buckling of fluid membranes in details.

Recently, the stress of torque tensors in fluid membranes were derived \cite{capo02,four07}.
When the membrane is curved, the mechanical surface tension is anisotropic
and deviated from the thermodynamic surface tension (energy to create a unit membrane area).
For a tubular membrane, the axial stress is finite while the azimuth stress is zero.

In this paper, we report on the shape and surface tension of buckled fluid membranes
using an analytical theory and numerical simulation.
Since the shape of buckled membrane can be analytically derived,
the anisotropy of the mechanical tensions can be investigated in details.
Buckling is one of the triggers for breaking membranes.
For example,
under shear flow, the formation of multi-lamellar vesicles
 \cite{diat93b,nett03} is considered to be induced by buckling instability \cite{zilm99}.
Shear suppression of the thermal undulations of the membrane and the
resulting reduction of the excess area induces buckling.
Our study revealed that buckling produces large anisotropy in the mechanical surface tension.
This anisotropy may play a role in the membrane stability.

\section{Simulation Method}

We employ a solvent-free meshless membrane model \cite{nogu06,nogu06a} 
to simulate buckling.
The fluid membrane is represented by a self-assembled one-layer sheet of particles.
The particles interact with each other via the potential
$U= \varepsilon( U_{\rm {rep}} +  U_{\rm {att}}) + k_{\rm {\alpha}} U_{\rm {\alpha}}$,
which consists of a repulsive soft-core potential $U_{\rm {rep}}$ with a 
diameter $\sigma$, an attractive potential $U_{\rm {att}}$, and a 
curvature potential $U_{\alpha}$.
The details of the model are given in Appendix.

We simulate single- and multi-layer
fluid membranes by Brownian dynamics 
in the $NVT$ ensemble with periodic boundary 
conditions in a rectangular box with side lengths $L_x$, $L_y$, and $L_z$.
The buckling is chosen in the $x$ direction.
The buckling occurs in the longest wave length $L_x$ in the $x$ direction.
When $L_x$ is gradually reduced,
the buckled membrane with large amplitude keep its direction along $x$ axis even at $L_x<L_y$.
The bending rigidity $\kappa_{\rm {cv}}$ of the membrane shows a linear dependence on the parameter $k_{\rm {\alpha}}$.
We calculate $\kappa_{\rm {cv}}$ from the thermal undulations of the planar membrane \cite{shib11}.
Three typical bending rigidities are chosen for the simulations: 
$\kappa_{\rm {cv}}/k_{\rm B}T= 9 \pm 0.2$, $21 \pm 0.5$, and $44 \pm 1$ for $k_{\rm {\alpha}}/k_{\rm B}T=5$, $10$, and $20$,
respectively,
where $k_{\rm B}T$ is the thermal energy.
The surface tensions are calculated
using $\gamma_x=-P_{xx} L_z$ and $\gamma_y=-P_{yy} L_z$, 
with the diagonal components of 
the pressure tensor $P_{\alpha\alpha} = (Nk_{\rm B}T - 
 \sum_{i} \alpha_{i}\frac{\partial U}{\partial {\alpha}_{i}} )/V$ 
for $\alpha \in \{x,y,z\}$,
since $P_{zz}\simeq 0$ for solvent-free models.
The numerical errors are estimated from three or more independent runs.

\section{Single buckled membrane}

\begin{figure}
\includegraphics[width=8.5cm]{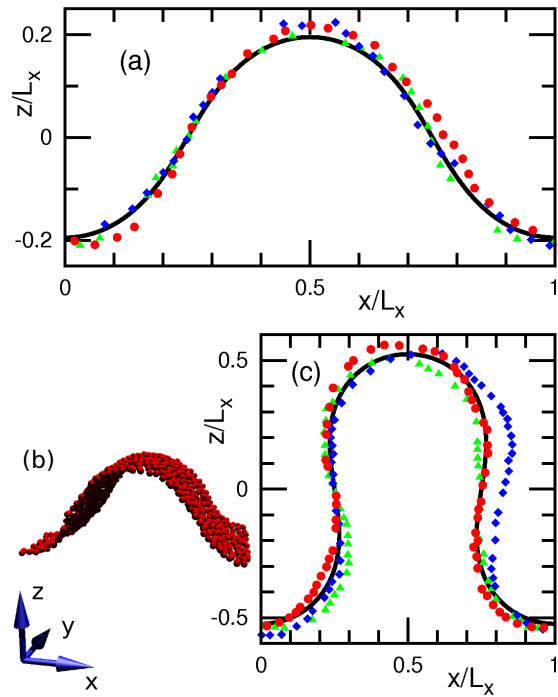}
\caption{\label{fig:snap}
(Color online)
Sliced shape of buckled fluid membranes at $L_x=35\sigma$  and $\kappa_{\rm {cv}}=21k_{\rm B}T$.
(a) $A_{xy}/A=0.75$, $L_y=24\sigma$,  and $N=800$.
(c) $A_{xy}/A=0.375$, $L_y=48\sigma$, and $N=3200$.
The solid curves are given by  Eqs. (\ref{eq:bkx}) and (\ref{eq:bky}).
The closed symbols ($\bullet, \triangle, \diamond$) represent 
membrane particles in the simulation
for three sequential sliced snapshots with time interval $2000\tau$.
The curve and points are shifted in the $x$ direction to overlap at $(x,z)=(0.25L_x,0)$.
(b) The 3D image of the simulated membrane shown with closed circles in (a).
}
\end{figure}

First, we consider the buckling of an isolated planar membrane.
Figure \ref{fig:snap} shows
buckled membranes at small projected areas $A_{xy}=L_xL_y$.
When the thermal undulations are neglected,
the shape of the membrane is given by the energy minimum of
the bending energy $F_{\rm {cv}}$ of the membrane with area constraints.
For the buckled membrane, it is given by
\begin{equation} \label{eq:fcv}
F_{\rm {cv}}=L_y\int_0^L \frac{\kappa_{\rm {cv}}}{2} \Big(\frac{d\theta}{ds}\Big)^2 ds
\end{equation}
with constant intrinsic area $A=L L_y$,
where $s$ and $\theta$ are the arc length and the tangential angle on $xz$ plane, respectively.
When we consider the membrane compressed by a constant force $\lambda L_y$ in the $x$ direction,
it is  an elastica problem
to minimize $F_{\rm {total}}= F_{\rm {cv}}+ \lambda L_x L_y$.
Also, the force $\lambda L_y$ can be considered as a Lagrange multiplier to fix the length $L_x$.
Euler's equation gives the shape equation
\begin{equation} \label{eq:shapeq1}
\frac{b^2}{2} \Big(\frac{d\theta}{ds}\Big)^2 =  \cos(\theta) - \cos(\theta_{\rm {max}}),
\end{equation}
where the characteristic length $b=\sqrt{\kappa_{\rm {cv}}/\lambda}$
and $\theta_{\rm {max}}$ is the maximum tangential angle \cite{true83,tsur86,toda}.
Then, the arc length $s$ is written as
$s = b \ F(\varphi,k)$, with $k=\sin(\theta_{\rm {max}}/2)$ and $\sin(\theta/2)=k \sin(\varphi)$,
where $F(\varphi,k)$ is the elliptic integral of the first kind with the elliptic modulus $0 \le k \le 1$ \cite{byrd54}.
The modulus $k$ is determined by the total arc length $L$ from 
\begin{equation} \label{eq:L}
\frac{L}{4b}=K(k), 
\end{equation}
where $K(k)=F(\pi/2,k)$ is the complete elliptic integral of the first kind.
The shape of the buckled membrane is expressed by
\begin{eqnarray}
x &=& 2b\ E({\rm am}(s/b,k),k)\ -\ s  \label{eq:bkx}, \\
z &=& 2 k b\ {\rm cn}(s/b,k), \label{eq:bky}
\end{eqnarray}
where $E(a,k)$, ${\rm am}(a,k)$, and ${\rm cn}(a,k)$ are the elliptic integral of the second kind,
Jacobi amplitude, and Jacobi elliptic function, respectively \cite{byrd54}.
Equations (\ref{eq:bkx}) and (\ref{eq:bky}) reproduce the buckled shape of the simulation very well
(see Fig. \ref{fig:snap}).
The thermal fluctuations give small undulations of the simulated 
membrane around the energy minimum shape (solid curve).
Strongly buckled membrane with $\theta_{\rm {max}}>\pi/2$ ($k^2>1/2$) 
has a $\Omega$ shape as shown in Fig. \ref{fig:snap}(c).
It is called class 4 of Euler's elastica \cite{true83}.

Next, we derive the surface tension for fixed $L_x$ and $L_y$ using elliptic functions.
We only consider the mechanical surface tension here.
The modulus $k$ is determined from the length ratio,
\begin{equation} \label{eq:Lx}
\frac{L_x}{L} = \frac{2 E(k)}{K(k)} - 1.
\end{equation}
The stress $\lambda$ is a variable given by $\lambda=16\kappa_{\rm {cv}} (K(k)/L)^2$.
The bending energy of the buckled membrane is given by
\begin{eqnarray}
F_{\rm {cv}} &=& \lambda \Big\{ (2k^2-1)L +L_x \Big\} \nonumber \\  \label{eq:fcvbk}
&=& \frac{32\kappa_{\rm {cv}} L_y}{L} \Big\{ (k^2-1)K(k)^2 -E(k)K(k) \Big\}.
\end{eqnarray}
The surface tensions in the $x$ and $y$ directions are given by
\begin{eqnarray} \label{eq:gx1}
\gamma_x &=& \frac{\ \partial F_{\rm {cv}}}{L_y \partial L_x}\Big|_{L_y} = - \lambda, \\ \label{eq:gy1}
\gamma_y &=& \frac{\ \partial F_{\rm {cv}}}{L_x \partial L_y}\Big|_{L_x} = - \lambda + \frac{2F_{\rm {cv}}}{L_xL_y},
\end{eqnarray}
respectively.
The  surface tension $\gamma_x$ is balanced by the compressed stress $\lambda$ as expected.
However, the surface tension $\gamma_y$ has the additional term $2F_{\rm {cv}}/L_xL_y$.
This anisotropy can be also derived from the stress tensor derived in Ref. \cite{capo02,four07}.
In contrast, solid sheets show much weaker correlation between the $x$ and $y$ directions 
because of the shear elasticity.

The buckling transition points are obtained from 
the condition required to satisfy Eq. (\ref{eq:L}).
Since $K(k)>K(0)=\pi/2$, 
it is written as $\lambda > \lambda_0=(2\pi/L)^2\kappa_{\rm {cv}}$ \cite{true83,toda}.
Therefore, the planar membrane becomes unstable at $\gamma = -(2\pi/L_x)^2\kappa_{\rm {cv}}$.
This result is in agreement with that estimated
from the instability of the lowest Fourier mode of the thermal undulations \cite{miln89,otte05,nogu06}.
This coincidence is not surprising because
the elliptic function reduces to a trigonometric function at $k=0$.

\begin{figure}
\includegraphics{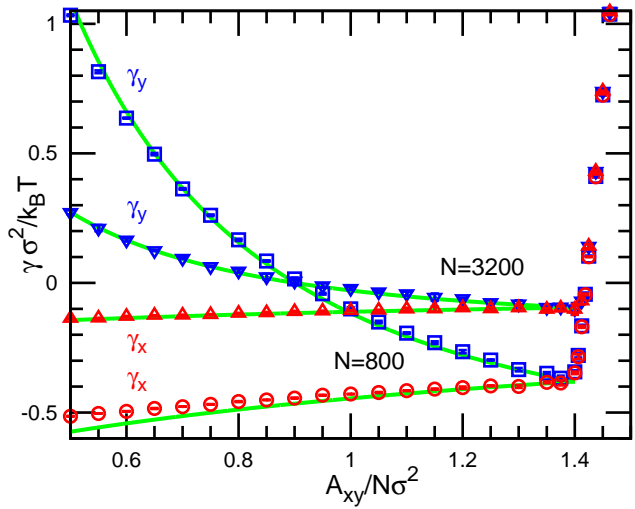}
\caption{\label{fig:gk10}
(Color online)
Area $A_{xy}$ dependence of the surface tension for $\kappa_{\rm {cv}}=21k_{\rm B}T$ 
and $L_y=24\sigma$ ($N=800$) or $L_y=48\sigma$ ($N=3200$).
The symbols ($\circ, \triangle$) and ($\Box$, $\triangledown$) 
represent the surface tension $\gamma_x$ and $\gamma_y$ in the simulation, respectively.
The solid lines are given by Eqs. (\ref{eq:gx1}) and  (\ref{eq:gy1}).
}
\end{figure}

Figure \ref{fig:gk10} shows the surface tension dependence on $A_{xy}$ for constant $L_y$.
At $A_{xy}<A_{xy}^{\rm c}=1.4 N\sigma^2$, the membrane is buckled,
and then the two surface tensions $\gamma_x$ and $\gamma_y$ show different values.
As $A_{xy}$ decreases, 
it is found that $\gamma_y$ increases,
while  $\gamma_x$ shows a gradual decrease.
Interestingly, the area decrease generates reduction of the counter stress in the $y$ direction.
It is surprising that $\gamma_y$ becomes positive at $A_{xy}/N\sigma^2<0.9$
so that the buckled membrane prefers to shrink in the $y$ direction.
The average surface tension $\gamma_{\rm {av}}=(\gamma_x+\gamma_y)/2$ also increases.
These simulation results are in good agreement with our theoretical prediction 
given by Eqs. (\ref{eq:gx1}) and (\ref{eq:gy1}) with the area $A=A_{xy}^{\rm c}$.
The larger membrane starts buckling at smaller $\lambda=-\gamma$,
and it has smaller $A_{xy}$ dependence of $\gamma_x$ and $\gamma_y$, 
since $\lambda_0 \propto L_x^{-2}$ (compare data for $N=800$ and $3200$ in Fig. \ref{fig:gk10}). 

\begin{figure}
\includegraphics{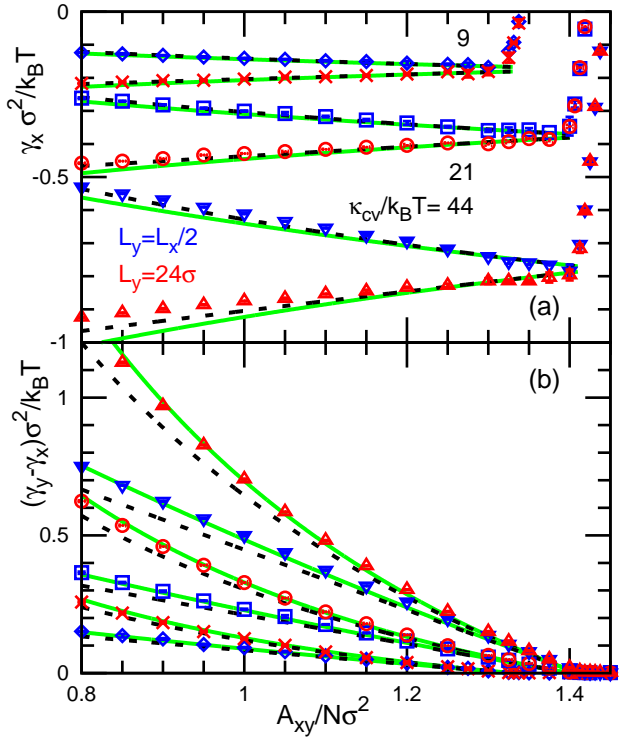}
\caption{\label{fig:g800}
(Color online)
Area $A_{xy}$ dependence of the surface tension for $\kappa_{\rm {cv}}/k_{\rm B}T=9$, $21$, or $44$ and $N=800$. 
The length $L_x$ is varied with constant $L_y=24\sigma$ ($\times,\circ,\triangle$) or constant ratio $L_y=L_x/2$ 
($\diamond, \Box, \triangledown$).
The solid lines are given by  Eqs. (\ref{eq:gx1}) and (\ref{eq:gy1}).
The dashed lines are given by the approximation,  Eqs. (\ref{eq:gx2}) and (\ref{eq:gy2}).
}
\end{figure}

When the aspect ratio $L_y/L_x$ is fixed,
the surface tensions show a different type of $A_{xy}$ dependence (see Fig. \ref{fig:g800}).
With decreasing $A_{xy}$,
$\gamma_x$ gradually increases, in contrast to its behavior for constant $L_y$,
and the tension difference $\gamma_y-\gamma_x$ increases weakly.
These effects are due to an increase in the arc length $L=A/L_y$ for the fixed aspect ratio.
Thus, the surface tensions are dependent on the projected area as well as 
the aspect ratio.

When the aspect ratio $L_y/L_x$ is allowed to change freely for a fixed projected area $A_{xy}=L_xL_y$,
the membrane elongates in the $x$ direction ($L_x \to \infty$)
in order to reduce the membrane bending energy ($L_y/L \to 0$ with constant $k$ in Eq. (\ref{eq:fcvbk}) ):
i.e. Mechanically, the membrane pushes the wall in the buckled direction more than 
that in the other direction.
This is a qualitative explanation of the anisotropic surface tension.
Thus, the buckling gives the effective shear elasticity to the membrane.

For small buckling amplitude at $1-A_{xy}/A_{xy}^{\rm c} \ll 1$,
the surface tensions can be expressed by polynomials.
Eq. (\ref{eq:Lx}) can be expanded to
$L_x/L =  1 - k^2 - k^4/8 + o(k^6)$  for small $k$.
From this relation and the expansions of 
Eqs. (\ref{eq:gx1}), and (\ref{eq:gy1}),
the surface tensions are expressed as
\begin{eqnarray} \label{eq:gx2}
\gamma_x&=& - \frac{\pi^2 \kappa_{\rm {cv}}}{4L^2}\Big( 5-\frac{L_x}{L}\Big)^2, \\ \label{eq:gy2}
\gamma_y-\gamma_x &=& \frac{8\pi^2 \kappa_{\rm {cv}}}{L_x L}\Big( 1-\frac{L_x}{L}\Big).
\end{eqnarray}
The area dependence can be clearly captured by these equations.
The surface tensions are determined by three quantities, 
$L_x/L=A_{xy}/A_{xy}^{\rm c}$, $A_{xy}^{\rm c}$, and $\kappa_{\rm {cv}}$.
These equations give a good approximation for $L_x/L \gtrsim 0.6$ 
(compare solid and dashed lines in Fig. \ref{fig:g800}).

Anisotropic surface tensions are also generated in a tubular membrane \cite{bo89,harm06,four07a}.
The surface tension in the axial direction is given by $\gamma_{\rm {ax}}=\kappa_{\rm {cv}}/R^2$,
where $R$ is the radius of the membrane tube;
while the surface tension is zero in the azimuth direction 
(The average surface tension $\gamma_{\rm {av}}=\gamma_{\rm {ax}}/2$).
The bending rigidity $\kappa_{\rm {cv}}$ of the membrane has been
measured using this axial tension in experiments \cite{bo89} 
and simulations \cite{harm06}.
Similarly, $\kappa_{\rm {cv}}$ can be measured from the surface tension of the buckled membrane.
We fit the curves in Fig. \ref{fig:g800}(b) by Eq. (\ref{eq:gy2}) 
with the fit parameters $\kappa_{\rm {cv}}$ and $A_{xy}^{\rm c}$ for $0.05<(A_{xy}^{\rm c}-A_{xy})/N\sigma^2<0.3$. 
This gives $\kappa_{\rm {cv}}/k_{\rm B}T=9.4 \pm 0.3$ ($9.3 \pm 0.3$), $22.3 \pm 0.1$ ($22.9 \pm 0.2$), and $47.4 \pm 0.1$ ($48.4 \pm 0.2$) for constant $L_y$ (constant ratio $L_y/L_x$)
 at $k_{\rm {\alpha}}/k_{\rm B}T=5$, $10$, and $20$, respectively.
These values are in reasonable agreement with the bending rigidities estimated from the thermal undulations
 of the planar membranes.
Compared to the tubular membrane, this simulation method
 is easy to apply to explicit solvent systems and bilayer membranes with low flip-flop frequency.
In the buckling,
the area difference between the upper and the lower leaflets of bilayers
are not changed, since the membrane deformation is symmetric.
The solvent is not enclosed by the membrane, so the solvent volume is conserved when the volume of the simulation box is fixed.
In contrast, a radius variation of the tubular membrane accompanies changes in the tube volume
and the area difference between the two leaflets.
Therefore, it has to take the Laplace pressure into account or requires 
an additional numerical technique to exchange the solvent particles or lipids between
the upper and lower sides of the bilayers.
The new buckling method is suitable for measuring the bending rigidity of these membranes.

\begin{figure}
\includegraphics{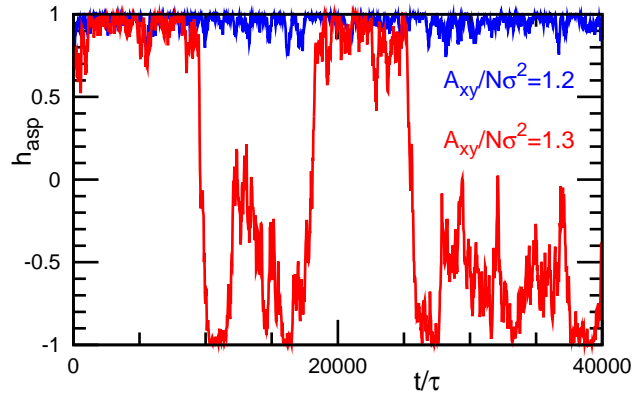}
\caption{\label{fig:flip}
(Color online)
Time development of
the aspect ratio $h_{\rm {asp}}$
of the buckled amplitude for $A_{xy}/N\sigma^2 = 1.2$ and $1.3$
at $\kappa_{\rm {cv}}/k_{\rm B}T=21$, $L_x=L_y$, and  $N=1600$. 
}
\end{figure}

We neglect the effects of thermal fluctuations in our analysis.
The excess area induced by the thermal undulations
shows a slight increase with deceasing area in the buckling simulation.
However, this does not have a large effect on the surface tensions.
For a squared membrane $L_x=L_y$,
the thermal fluctuations can induce a flip between the buckling in the $x$ and $y$ directions
 slightly below the buckling transition point.
In the meshless membrane at $L_x=L_y$ and  $N=1600$, 
this flip is observed for $A_{xy}/N\sigma^2 \gtrsim 1.3$.
Figure \ref{fig:flip} shows the time development of
the aspect ratio
of the buckled amplitude, 
\begin{equation} \label{eq:asp}
h_{\rm {asp}}= \frac{|h(q_{x1})|^2-|h(q_{y1})|^2}{|h(q_{x1})|^2+|h(q_{y1})|^2},
\end{equation}
where $h(q_{x1})=\sum_j z_j \exp(-2\pi i  x_j/L_x)$
and $h(q_{y1})=\sum_j z_j \exp(-2\pi i y_j/L_y)$.
The membrane changes the buckling direction
at $A_{xy}/N\sigma^2 = 1.3$,
while not at  $A_{xy}/N\sigma^2 = 1.2$.

\begin{figure}
\includegraphics[width=8.5cm]{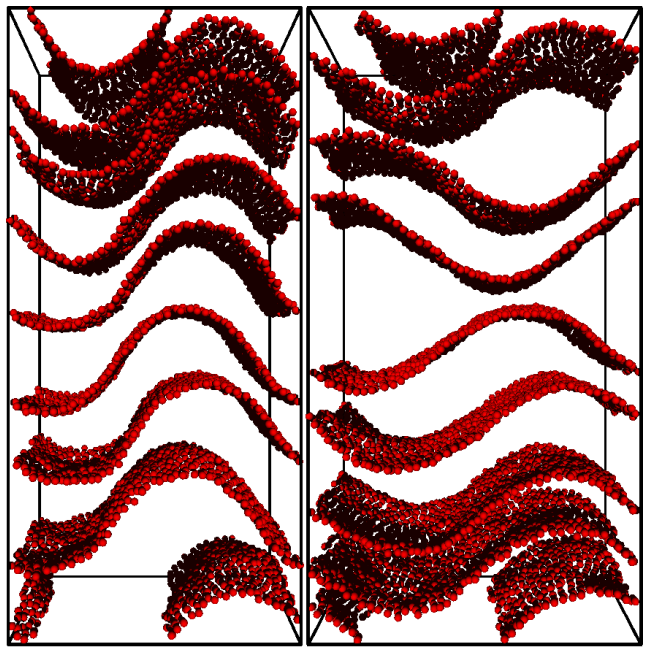}
\caption{\label{fig:mul1}
(Color online)
Snapshots of multiple buckled membranes at $N=6400$, $N_{\rm {mb}}=8$, $L_y/\sigma=24$,
$\kappa_{\rm {cv}}/k_{\rm B}T=21$, and $d_{\rm m}/\sigma=10$.
Left and right panels show the membranes with
complete and partial phase synchronization at 
 $A_{xy}N_{\rm {mb}}/N\sigma^2 = 1.1$ and $1.25$, respectively.
}
\end{figure}

\section{Multiple buckled membranes}

\begin{figure}
\includegraphics{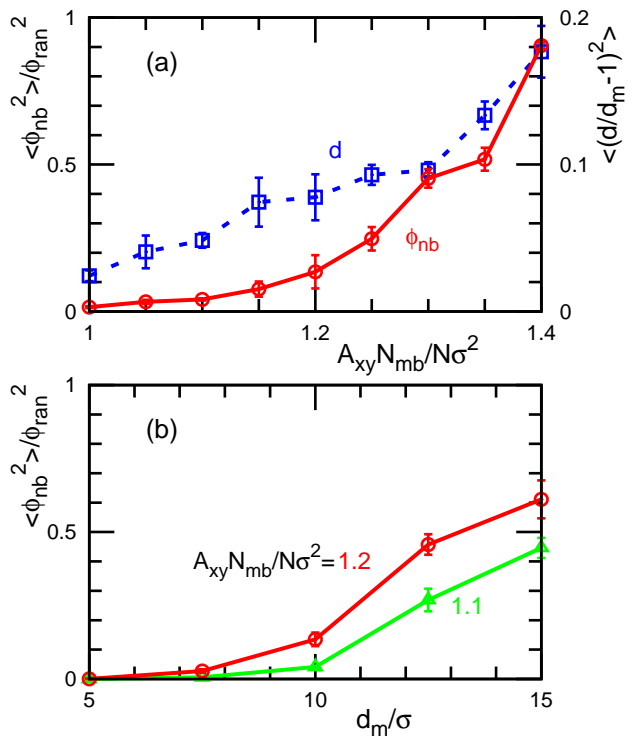}
\caption{\label{fig:mul2}
(Color online)
Phase synchronization of buckled membranes at $N=6400$, $N_{\rm {mb}}=8$, $L_y/\sigma=24$,
and $\kappa_{\rm {cv}}/k_{\rm B}T=21$.
(a) The area $A_{xy}$ dependence of
the fluctuation amplitudes of the distance $\langle (d/d_{\rm m}-1)^2 \rangle$
($\Box$)
and the phase difference $\langle \phi_{\rm {nb}}^2 \rangle$ ($\circ$)  between neighboring membranes at $d_{\rm m}/\sigma=10$.
(b) The phase difference $\langle \phi_{\rm {nb}}^2 \rangle$ dependence on the mean distance $d_{\rm m}$ between neighboring membranes at  $A_{xy}N_{\rm {mb}}/N\sigma^2 = 1.1$ ($\triangle$) and $1.2$ ($\circ$).
}
\end{figure}

The membrane thermal undulations
generate entropic repulsive force $f \propto d^{-3}$
between tensionless fluid membranes with neighboring membrane distance $d$ \cite{helf73}.
Here, we consider the interactions between the buckled membranes.
Since the buckled membranes have greater height amplitudes than the tensionless membrane,
the membranes have stronger short-range repulsion with decreasing area $A_{xy}$ (see Fig. \ref{fig:mul1}).
Then, the fluctuation amplitude of the neighboring membrane distance $\langle (d/d_{\rm m}-1)^2 \rangle$
decreases for the fixed mean distance $d_{\rm m}=L_Z/N_{\rm {mb}}$,
where $N_{\rm {mb}}$ is the number of the membranes (see the dashed line in Fig. \ref{fig:mul2}(a)).

Along with this reduction in the distance fluctuation,
it is found that the buckling of the neighboring membrane becomes synchronized in phase.
The phase $\phi$ is calculated
from the lowest Fourier mode of the membrane height, 
$h(q_{x1})=h_{\rm {am}}\exp(-i \phi)=\sum_j z_j \exp(-2\pi i  x_j/L_x)$,
for each membrane.
The phase difference $\phi_{\rm {nb}}$ between neighboring membranes
approaches zero as $A_{xy}$ decreases.
The phase deviation of the neighboring membrane  $\langle \phi_{\rm {nb}}^2 \rangle/\phi_{\rm {ran}}^2$
is shown as a solid line in Fig. \ref{fig:mul2}(a),
where $\phi_{\rm {nb}}^2$
is normalized by the average for the random distribution,
 $\langle \phi_{\rm {nb}}^2 \rangle= \phi_{\rm {ran}}^2=\pi^2/3$.
Thus, the translational order of the buckled shape appears in the $z$ direction.
This ordering is not a discrete transition but a gradual change,
since it is a quasi-one-dimensional system.
As  $A_{xy}$ decreases,
clusters of the synchronized membrane appear,
and then all of membranes are synchronized at $A_{xy}N_{\rm {mb}}/N\sigma^2 \leq 1.1$.
(see the snapshots in Fig. \ref{fig:mul1} and movies in EPAPS \cite{epaps1}).
The interactions between the membranes little change their surface tension
in the simulated area range.
As the mean distance $d_{\rm m}$ increases, the synchronization requires smaller $A_{xy}$ 
(see Fig. \ref{fig:mul2}(b)).
This synchronized buckling may act as a nucleus to form multi-lamellar vesicles in shear flow.

\section{Summary}

We studied the elastica of fluid membrane.
The buckled shape and surface tension parallel to the buckling ($x$) direction
are expressed by the formula used for the elastica of solid sheets.
However,  unlike the solid sheets,
the surface tension of the fluid membranes
in the perpendicular ($y$) direction
shows large increases for decreasing projected area $A_{xy}$.
Additionally, multi-lamellar buckled membranes
were found to have phase synchronization.

The anisotropic surface tension would also appear for the buckled Langmuir monolayers
in the fluid phase.
It can be experimentally checked, if one separately measures the stress in two lateral directions.
These anisotropy and synchronization may play a role in the breakup of 
the membrane under external fields. Recent experiments show that
collagen-containing tubular vesicles exhibit elastica-shape under magnetic field \cite{suzu07}.
It will be interesting to investigate the coupling of mechanical and external-field induced bucklings.

\begin{acknowledgments}
We would like to thank T. Nakamura and W. Shinda (AIST) for stimulating discussion.
This work is supported by KAKENHI (21740308) from
the Ministry of Education, Culture, Sports, Science, and Technology of Japan.
\end{acknowledgments}

\begin{appendix}
\section{Details of simulation model and method}

A membrane consists of $N$ particles, which possess no internal degrees 
of freedom.  The particles interact with each other via a potential
\begin{equation} \label{eq:potential}
U= \varepsilon( U_{\rm {rep}} +  U_{\rm {att}}) 
                      + k_{\rm {\alpha}} U_{\rm {\alpha}},
\end{equation} 
which consists of a repulsive soft-core potential $U_{\rm {rep}}$ with a 
diameter $\sigma$, an attractive potential $U_{\rm {att}}$, and a 
curvature potential $U_{\alpha}$.  All three potentials only depend on 
the positions ${\bf r}_i$ of the particles.
In this paper, we employ the curvature potential
based on the first-order moving least-squares (MLS) method 
(model II in Ref.~\onlinecite{nogu06}).
We briefly outline here the simulation technique,
since the membrane model is explained in more detail in 
Ref.~\onlinecite{nogu06}.

\subsection{Curvature Potential \label{sec:cv_pot}}

A Gaussian function with $C^{\infty}$ cutoff~\cite{nogu06} 
is employed as a weight function 
\begin{equation} \label{eq:wmls}
w_{\rm {mls}}(r_{i,j})=\left\{ 
\begin{array}{ll}
\exp (\frac{(r_{i,j}/r_{\rm {ga}})^2}{(r_{i,j}/r_{\rm {cc}})^n -1})
& (r_{i,j} < r_{\rm {cc}}) \\
0  & (r_{i,j} \ge r_{\rm {cc}}) 
\end{array}
\right.
\end{equation}
where $r_{i,j}$ is the distance between particles $i$ and $j$.
This function is smoothly cut off at $r_{i,j}=r_{\rm {cc}}$.
We use here the parameters $n=12$, $r_{\rm {cc}}=3\sigma$, 
and $r_{\rm {ga}}=1.5\sigma$. 

The degree of deviation from a plane, ``aplanarity'' is defined as
\begin{eqnarray}
\alpha_{\rm {pl}} &=& \frac{9D_{\rm {w}}}{T_{\rm {w}}M_{\rm {w}}} \\
 &=& \frac{9\lambda_1\lambda_2\lambda_3} {(\lambda_1+\lambda_2+\lambda_3)
    (\lambda_1\lambda_2+\lambda_2\lambda_3+\lambda_3\lambda_1)},
                                                \nonumber
\end{eqnarray} 
where $\lambda_1$, $\lambda_2$, and $\lambda_3$ are
the eigenvalues of the weighted gyration tensor,
$a_{\alpha\beta}= \sum_j (\alpha_{j}-\alpha_{\rm G})
(\beta_{j}-\beta_{\rm G})w_{\rm {mls}}(r_{i,j})$,
where $\alpha, \beta=x,y,z$ and ${\bf r}_{\rm G}=\sum_j {\bf r}_{j}w_{\rm {mls}}(r_{i,j})/\sum_j w_{\rm {mls}}(r_{i,j})$.
The aplanarity can be calculated from three invariants of the tensor:
 $D_{\rm {w}}$ and $T_{\rm {w}}$ are determinant and trace, 
respectively, and $M_{\rm {w}}$ is the sum of its three minors, 
$M_{\rm {w}}= a_{xx}a_{yy}+a_{yy}a_{zz}+a_{zz}a_{xx}
 -a_{xy}^2-a_{yz}^2-a_{zx}^2$.

The aplanarity $\alpha_{\rm {pl}}$ takes values in the interval 
$[0, 1]$ and represents the degree of deviation from a plane.
This quantity acts like $\lambda_1$ for 
$\lambda_1 \ll \lambda_2, \lambda_3$, since 
$\alpha_{\rm {pl}}\simeq 9\lambda_1/(\lambda_2+\lambda_3)$ in this
limit. Therefore, we employ the curvature potential
\begin{eqnarray}
  U_{\rm {\alpha}}= \sum_i \alpha_{\rm {pl}}({\bf r}_{i}),
\label{eq:ual}
\end{eqnarray} 
where $\alpha_{\rm {pl}}({\bf r}_{i})=0$ when the $i$-th particle 
has two or less particles within the cutoff distance 
$r_{i,j}<r_{\rm {cc}}$.  This potential increases with increasing 
deviation of the shape of the neighborhood of a particle from a 
plane, and favors the formation of quasi-two-dimensional membrane
aggregates.

\subsection{Attractive and Repulsive Potentials \label{sec:2d_pot}}
  
The particles interact with each other in the quasi-two-dimensional 
membrane surface via the potentials $U_{\rm {rep}}$ and $U_{\rm {att}}$.
The particles have an excluded-volume interaction via the repulsive
potential 
\begin{eqnarray}
   U_{\rm {rep}}=\sum_{i<j} \exp(-20(r_{i,j}/\sigma-1)+B)
           f_{\rm {cut}}(r_{i,j}/\sigma),
\label{eq:rep}
\end{eqnarray}
with $B=0.126$, and a $C^{\infty}$-cutoff function \cite{nogu06} 
\begin{equation} \label{eq:cutoff}
f_{\rm {cut}}(s)=\left\{ 
\begin{array}{ll}
\exp\{A(1+\frac{1}{(|s|/s_{\rm {cut}})^n -1})\}
& (s < s_{\rm {cut}}) \\
0  & (s \ge s_{\rm {cut}}) 
\end{array}
\right.
\end{equation}
is employed.  
All orders of derivatives of $f_{\rm {cut}}(s)$ are continuous at 
the cutoff. In Eq.~(\ref{eq:rep}), we use the parameters
$n=12$, $A=1$, and $s_{\rm {cut}}=1.2$.

A solvent-free membrane
 requires an attractive interaction which mimics 
the "hydrophobic" interaction.  We employ a potential $U_{\rm {att}}$ 
of the local density of particles, 
\begin{eqnarray} \label{eq:dens}
\rho_i= \sum_{j \ne i} f_{\rm {cut}}(r_{i,j}/\sigma),
\end{eqnarray} 
with the parameters $n=12$, $s_{\rm {half}}=1.8$, and $s_{\rm {cut}}=2.1$. 
The factor $A$ in Eq. (\ref{eq:dens}) is determined such that 
$f_{\rm {cut}}(s_{\rm {half}})=0.5$, which implies
$A=\ln(2) \{(s_{\rm {cut}}/s_{\rm {half}})^n-1\}$.
Here, $\rho_{i}$ is the number of particles in a 
sphere whose radius is approximately 
$r_{\rm {att}} = s_{\rm {half}}\sigma$. The potential $U_{\rm {att}}$ 
is given by
\begin{eqnarray}
U_{\rm {att}} = \sum_{i} 0.25\ln[1+\exp\{-4(\rho_i-\rho^*)\}]- C,
\end{eqnarray} 
where $C= 0.25\ln\{1+\exp(4\rho^*)\}$.
For $\rho_i<\rho^*$, the potential is approximately 
$U_{\rm {att}}\simeq -\rho_i$ and therefore acts like a pair potential 
with $U_{\rm {att}}\simeq -\sum_{i<j} 2 f_{\rm {cut}}(r_{i,j}/\sigma)$.
For $\rho_i>\rho^*$, this function saturates to the constant $-C$.
Thus, it is a pairwise potential with cutoff at densities higher 
than $\rho_i>\rho^*$.
In this paper, we use  $\varepsilon/k_{\rm B}T=4$ and $\rho^*=6$.

\subsection{Dynamics}

The buckling of the membrane is simulated by Brownian dynamics
(molecular dynamics with Langevin thermostat).
 The motion 
of particles is determined by the underdamped Langevin equations
\begin{equation}
m \frac{d^2 {\bf r}_{i}}{dt^2}=
- \frac{\partial U}{\partial {\bf r}_{i}}
- \zeta \frac{d {\bf r}_{i}}{dt} + {\bf g}_{i}(t), 
\end{equation} 
where $m$ is the mass of a particle and $\zeta$ the friction
constant. ${\bf g}_{i}(t)$ 
is a Gaussian white noise which obeys the fluctuation-dissipation theorem.
We employ the time unit $\tau=\zeta\sigma^2/k_{\rm B}T$
with $m= \zeta\tau$.
The Langevin equations are integrated by the leapfrog algorithm 
\cite{alle87,nogu11} with a time step of $\Delta t=0.005\tau$.

\end{appendix}

\end{document}